\documentclass[superscriptaddress,showpacs,floatfix,twocolumn,prl,reprint]{revtex4-1}

\usepackage[T1]{fontenc}
\usepackage[english]{babel}
\usepackage{amsmath,amsfonts,amsthm,amssymb}
\usepackage{mathrsfs}
\usepackage{graphicx}
\DeclareMathAlphabet{\mathpzc}{OT1}{pzc}{m}{it}

\begin{document}
\title{Spontaneous chiral symmetry breaking in model bacterial suspensions}

\author{Rebekka E.\ Breier}
\affiliation{Max Planck Institute for Dynamics and Self-Organization (MPIDS), Am Fassberg 17,
37077 G\"{o}ttingen, Germany}
\author{Robin L.\ B.\ Selinger}
\affiliation{Chemical Physics Interdisciplinary Program, Liquid Crystal Institute, Kent State University, Kent, OH, USA}
\author{Giovanni Ciccotti}
\affiliation{Department of Physics, University of Rome ``La Sapienza'', P.le A. Moro 5, 00185 Rome, Italy}
\affiliation{School of Physics, University College Dublin, Belfield, Dublin 4, Ireland}
\author{Stephan Herminghaus}
\affiliation{Max Planck Institute for Dynamics and Self-Organization (MPIDS), Am Fassberg 17,
37077 G\"{o}ttingen, Germany}
\author{Marco G.\ Mazza}
\affiliation{Max Planck Institute for Dynamics and Self-Organization (MPIDS), Am Fassberg 17,
37077 G\"{o}ttingen, Germany}

\date{\today}

\begin{abstract}
Chiral symmetry breaking is ubiquitous in biological systems, from DNA to bacterial suspensions. A key unresolved problem is how chiral structures may spontaneously emerge from achiral interactions. We study a simple model of bacterial suspensions in three dimensions that effectively incorporates active motion and hydrodynamic interactions. We perform large-scale molecular dynamics simulations (up to $10^6$ particles) and describe stable (or long-lived metastable) collective states that exhibit chiral organization although the interactions are achiral. We elucidate under which conditions these chiral states will emerge and grow to large scales. We also study a related equilibrium model that clarifies the role of orientational fluctuations.
\end{abstract}

\maketitle

The emergence of biological structures that spontaneously break chiral symmetry has no clear explanation yet \cite{norden_asymmetry_1978}. 
For example, amino acids naturally occur in living beings only as left-handed enantiomers while DNA occurs only as right-handed. Bacterial suspensions also show striking examples of chiral behavior \cite{ben-jacob_cooperative_1995,wioland_confinement_2013}.
Moreover, colonies of the amoeba {\it Dictyostelium discoideum} aggregate in swirling localized vortices \cite{nicol_cell-sorting_1999,levine_swarming_2006}.
In not living, simple systems spontaneous chiral symmetry breaking has also been observed, e.g., in smectic films or close-packed confined spheres \cite{selinger_chiral_1993,pickett_spontaneous_2000,edlund_chiral_2012}.

Although great experimental and theoretical efforts have been made to unravel the physical mechanisms governing the collective states of bacterial suspensions \cite{lauga_hydrodynamics_2009,ramaswamy_mechanics_2010,vicsek_collective_2012,marchetti_hydrodynamics_2013}, progress in this field is hindered by the fact that standard tools of statistical mechanics, such as the fluctuation-dissipation theorem, are not applicable, because the fluctuations are not directly coupled to external perturbations \cite{mizuno_nonequilibrium_2007,chen_fluctuations_2007}.
Here, we show that a simple model of self-propelled particles (SPP) with achiral interactions may exhibit a chiral symmetry breaking under certain boundary conditions. 
We investigate the conditions which favor the formation of a stable (or long-lived metastable) chiral pattern from an isotropic state. 
Finally, we relate our model of SPP to an equilibrium statistical physics model which can elucidate how general the formation of a  chiral pattern is.

We model bacterial suspensions through simple interactions among SPP that effectively mimic the active motion of bacteria. 
The SPP are modeled as point particles moving in the direction of their intrinsic orientation $\hat{e}_i$.
It is a common choice to choose a constant speed $v_0$ because at small Reynolds numbers rapid fluctuations of the speed are exponentially damped. 
Aiming for a simple model system, we choose a nematic interaction that has head-tail symmetry. Nematic interactions represent to leading order the effective hydrodynamic interactions among bacteria \cite{baskaran_statistical_2009}. The equations of motion for particle $i$ read
\begin{subequations}\label{eq:differential}
\begin{align}
\dot{\vec{r}}_i &= v_0 \hat{e}_i \, ,\\
\dot{\hat{e}}_i &= -\gamma \dfrac{\partial U}{\partial \hat{e}_i} + \vec{\xi}_i(t)
\label{eq:differential_orientations}
\end{align}
\end{subequations}
where $\vec{r}_i$ and $\hat{e}_i$ are the position and direction, respectively, of the $i$-th particle, such that $|\hat{e}_i|^2=1$ and $\hat{e}_i\cdot\dot{\hat{e}}_i=0$ at all times.
The noise $\vec{\xi}(t)$ is a uniformly distributed vector on the surface of a sphere of radius $\eta$ \cite{czirok_collective_1999}, and $U=-\frac{1}{2}\sum_{i=1}^N\frac{1}{n_i}\sum_{j\in n_i} \left(\hat{e}_i \cdot \hat{e}_j \right) ^2$ is the Lebwohl-Lasher potential that induces nematic alignment \cite{lebwohl_nematic-liquid-crystal_1972},
where the second sum extends to the $n_i$ neighbors of particle $i$ within a sphere of radius $\epsilon$. 
A similar potential has been studied in two spatial dimensions by a number of authors \citep{vicsek_novel_1995,gregoire_onset_2004,peruani_mean-field_2008,ginelli_large-scale_2010}, however, past work has focused on the limit of fast angular relaxation which leads to finite-difference equations. 
Instead, we explicitly solve Eq.~\eqref{eq:differential} leaving $\gamma$ as an explicit parameter. 

We perform molecular dynamics simulations of a three-dimensional system with up to $N=10^6$ particles in a cubic box of volume $V=L^3$. 
We employ an explicit Euler algorithm with time-step $\delta t=0.1$ for the translational and orientational dynamics, where to satisfy the constraints on $\hat{e}_i$ we use the algorithm of \cite{singer_thermodynamic_1977,fincham_more_1984,ilnytskyi_domain_2002}.
We implement an efficient neighbor search \cite{boris_vectorized_1986,weinketz_two-dimensional_1993} that allows us to reach large system sizes, and we apply the usual periodic boundary conditions in all three dimensions. 
All following results are shown for $N=66^3$, $v_0=0.5$, and $\gamma=0.1$. 
The interaction range sets the length scale in our system ($\epsilon=1$).

To quantify the degree of nematic alignment we consider the nematic order parameter $S$ defined as the largest eigenvalue of the nematic order tensor $\mathbf{Q}=\tfrac{1}{2N}\sum_{i=1}^N[3\hat{e}_i\otimes\hat{e}_i-\mathbf{I}]$ where $\otimes$ is the tensor product.
The nematic director $\hat{d}$ is defined as the eigenvector associated to $S$. 
The local director $\hat{d}^\text{\,loc}$ is accordingly defined by replacing all $N$ particles in the definition of $\mathbf{Q}$ with a subset of $N_\alpha$ particles (e.g., contained in a layer of finite thickness).

\begin{figure}
\includegraphics[width=\columnwidth]{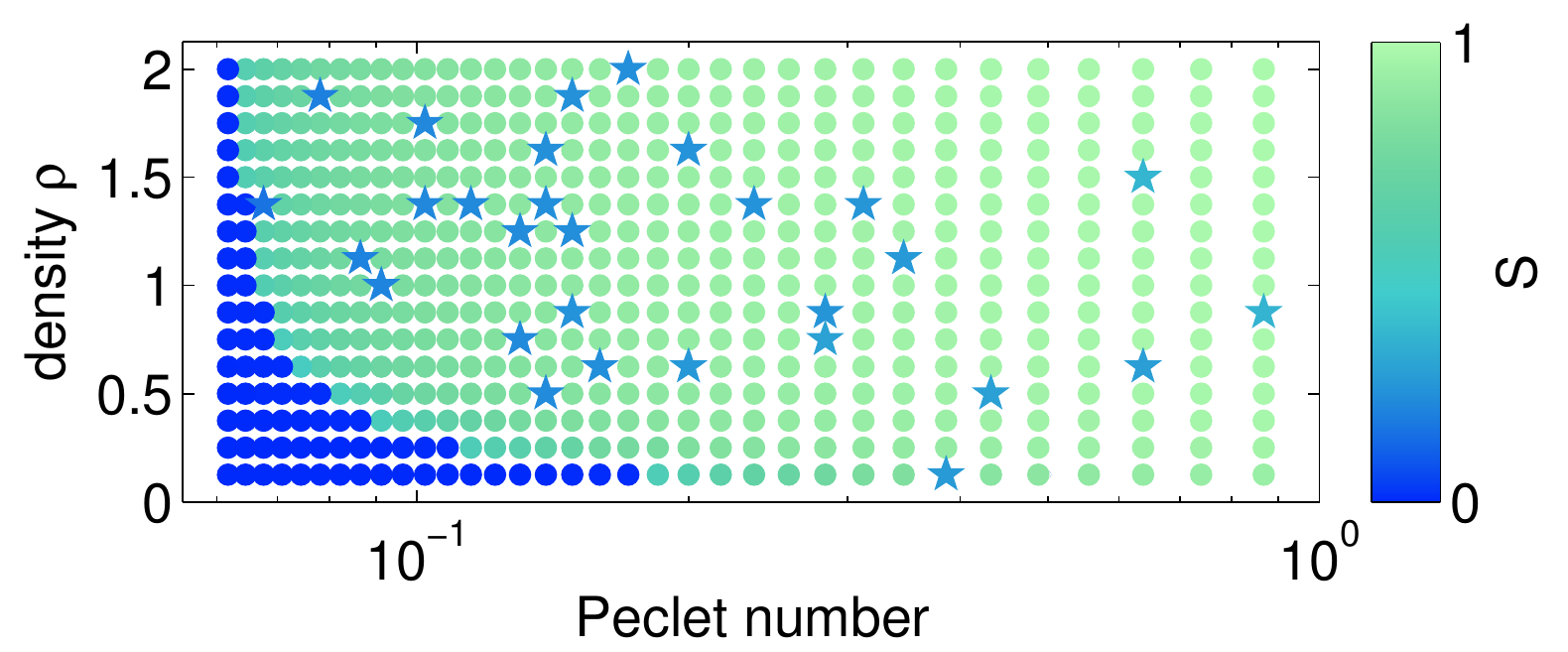}
\caption{(Color online) Phase diagram of the system of self-propelled particles in terms of P\'eclet number $\mathpzc{P}$ and density $\rho$. The color indicates the global nematic order parameter. The symbols indicate different phases: $\bigstar$ is a chiral pattern, while $\bullet$ are either nematic or isotropic patterns.}
\label{fig:phase_diagram}
\end{figure}
Figure~\ref{fig:phase_diagram} shows the nonequilibrium phase diagram of the system in terms of particle density $\rho=N/V$ and P\'eclet number
\begin{equation}
\mathpzc{P} \equiv \frac{\text{advection}}{\text{diffusion}} = \frac{\epsilon v_0}{\epsilon^2\eta^2/\gamma} = \frac{v_0\gamma}{\epsilon\eta^2}\,,
\end{equation}
which are useful to characterize bacterial or algal suspensions \cite{stocker_ecology_2012}.
At sufficiently large $\mathpzc{P}$, the system is in the nematic state independently of $\rho$ with $S\in[0.5,1]$. 
Moreover, as observed in \cite{ginelli_large-scale_2010} (but in 2D), the system's steady-state is spatially homogeneous with sub-populations moving in opposite directions. 
As $\mathpzc{P}$ decreases there is a clear transition to  a spatially homogeneous, isotropic state.

\begin{figure}
\centering
\includegraphics[width=\columnwidth]{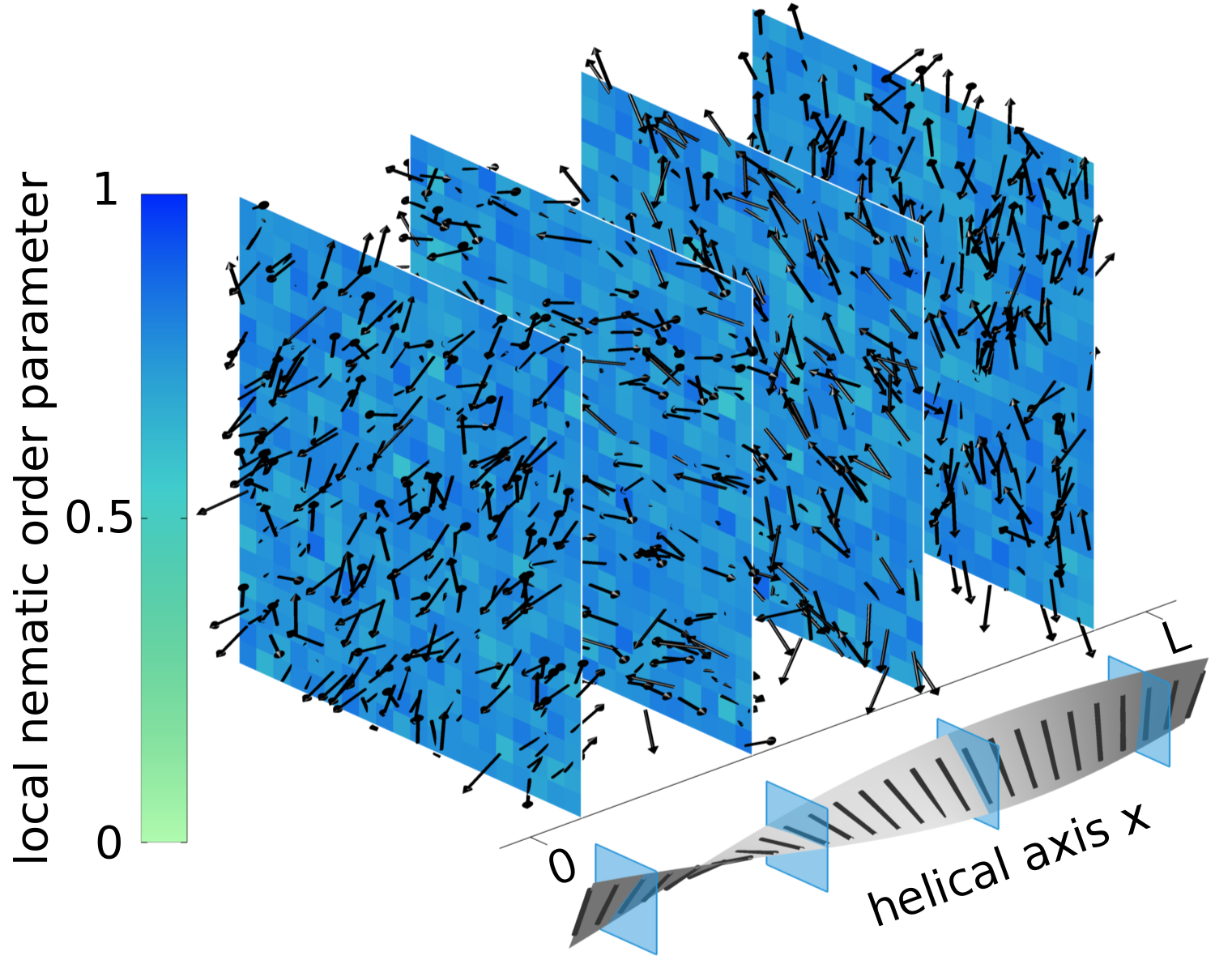}
\caption{(Color online) Cross-sections  of a typical chiral configuration  ($\rho=1.625, \mathpzc{P}=0.08$). The color code represents the local nematic order parameter and the arrows represent the SPP. The ribbon in the bottom shows the helicoidal behavior of $\hat{d}^\text{\,loc}$ in different cross-sections.}
\label{fig:snapshot}
\end{figure}

\begin{figure}
\centering
\includegraphics[width=\columnwidth]{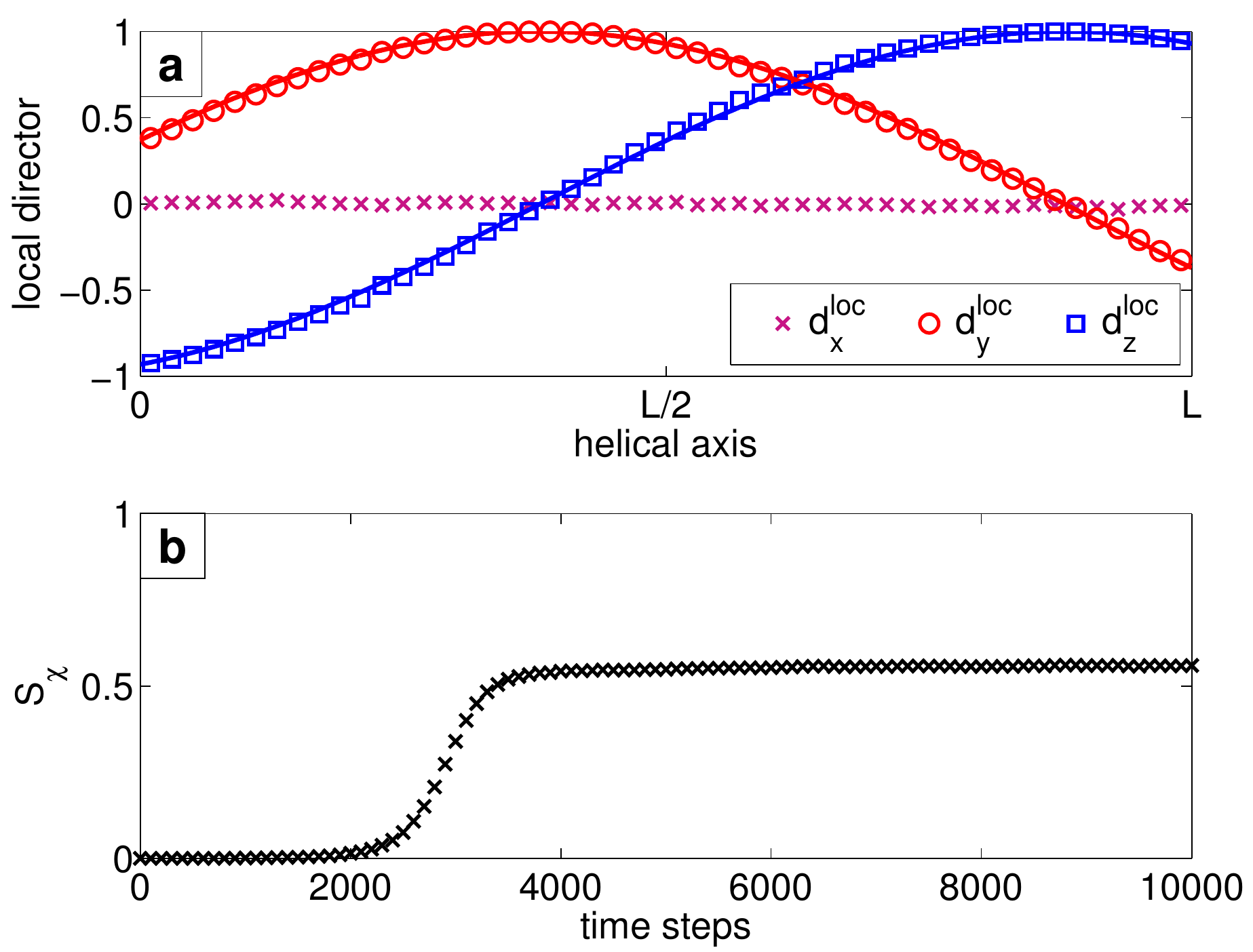}
\caption{(Color online) (a) Components of $\hat{d}^{\,\text{loc}}$ along the helical axis ($\hat{x}$) of the chiral pattern (see Fig.~\ref{fig:snapshot}). The symbols are the components of $\hat{d}^\text{\,loc}$ (in slices normal to the helical axis) and the solid lines are sinusoidal least square fits $d_{y,z}^\text{loc}=\cos\left(\pi x/L +\phi_{y,z}\right)$. (b) Time evolution of the chiral order parameter for the same simulation. The transient evolution before reaching the steady state can extend up to $10^4$ time steps}
\label{fig:director_components}
\end{figure}

In the region of the phase diagram where $\mathpzc{P}$ and $S$ are large, the system may also develop states where there is no single, global nematic director, but rather $\hat{d}^\text{\,loc}$ rotates in space forming a helical structure. 
Figure~\ref{fig:snapshot} shows four cross-sections of the system equally spaced along the helical axis. 
The particles contained in each of these cross-sections of width $\delta\approx1.1\epsilon$ still preserve nematic ordering with a well-defined $\hat{d}^\text{\,loc}$. 
As one moves along the helical axis this local nematic director slowly rotates with a constant twist angle. 
We show in Fig.~\ref{fig:director_components}(a) how the components of $\hat{d}^\text{\,loc}$ vary along the helical axis (conventionally called $x$-axis). 
The profiles of the $y$ and $z$ components of the director are very well fit by sinusoidal functions, as expected for helical structures. 
The axis of the helix is in most cases, like in this example, parallel to one of the box edges with the pitch being $2L$ due to the periodic boundary conditions together with the nematic interactions. 
However, the helix can also be found at an angle with the box edge (e.g., along one of the diagonals, with the pitch adjusted accordingly).
We observe left- and right-handed helices with equal probability, and, additionally, once a chiral state is formed it is stable (at least up to $2\cdot10^6$ time steps)\cite{Note_initial_conditions,Note_Mersenne}.

To characterize a chiral state a pseudoscalar order parameter is useful. 
The simplest combination of orientations and distances that provides a pseudoscalar is 
$\left(\hat{e}_i\times\hat{e}_j\right)\cdot\left(\vec{r}_i-\vec{r}_j\right)$ \cite{memmer_computer_2000}.
We define the chiral order parameter averaged over all particles as
\begin{equation}\label{eq:chiral_order_parameter}
S_\chi = -\frac{\pi^3}{6(4-\pi)} \frac{1}{N} \sum_{i=1}^{N} \frac{1}{N_i} \sum_{j=1}^{N_i}
\left[\left(\hat{e}_i \times \hat{e}_j \right) \cdot
\frac{\vec{r}_{ij}}{\left|\vec{r}_{ij}\right|}\right] \left(\hat{e}_i
\cdot \hat{e}_j \right)
\end{equation}
where $\vec{r}_{ij}\equiv \vec{r}_i-\vec{r}_j$ and the sum over $j$ includes all $N_i$ particles in a sphere centered on $\vec{r}_i$ with radius $L/4$ \cite{Note_normalization}.
$S_\chi$ is a pseudoscalar symmetric for $\hat{e}_i\rightarrow -\hat{e}_i$ and vanishes in both the nematic and the isotropic case. 
It is normalized so that $S_\chi=+1(-1)$ indicates a left-handed (right-handed) chiral structure.
Figure \ref{fig:director_components}(b) shows a typical evolution of $S_\chi$.


\begin{table}
\begin{ruledtabular}
\begin{tabular}{ccc}
$\mathpzc{P}$ & $P(\left|S_\chi\right|>0.2)$ & $\left<\left|S_\chi\right|\right>_\chi$ \\ \hline
0.08 & 3.3\% & 0.48\\
0.10 & 6.7\% & 0.64\\
0.14 & 5.7\% & 0.74\\
0.20 & 4.0\% & 0.82\\
\end{tabular}
\end{ruledtabular}
\caption{\label{tab:prob} Probability $P(\left|S_\chi\right|>0.2)$ of the formation of a chiral structure for different P\'{e}clet numbers (300 independent simulations each, $\rho=1$). $\left<\left|S_\chi\right|\right>_\chi$ denotes the mean of the absolute chiral order parameter conditional to the simulations resulting in a chiral structure with the helical axis parallel to the box edge.}
\end{table}

An estimate for the probability of the formation of a chiral pattern from 300 independent simulations for different values of $\mathpzc{P}$ is shown in Table~\ref{tab:prob}. 
Although the statistics is limited, this probability exhibits a maximum for $\mathpzc{P}=0.10$. 
$\left<\left|S_\chi\right|\right>_\chi$ increases as the P\'{e}clet number increases.
The reason is the same as for the increase of $S$ with increasing $\mathpzc{P}$ (Fig.~\ref{fig:phase_diagram}): Because the system has smaller orientational fluctuations for large $\mathpzc{P}$, both order parameters increase. 

\begin{figure}
\centering
\includegraphics[width=\columnwidth]{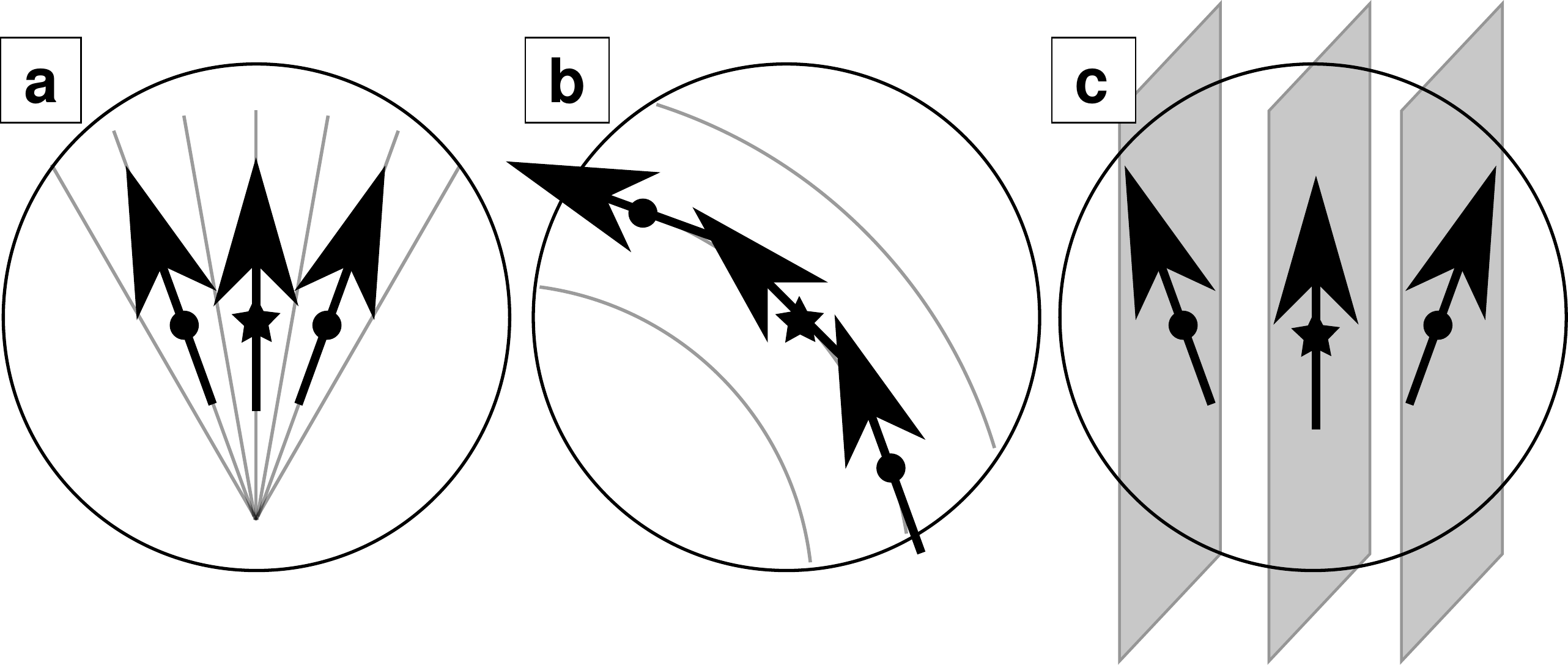}
\caption{Ideal splay (a), bend (b), and twist (c) deformations of the director field. The reference particle ($\bigstar$) does not experience any torque from the neighboring particles ($\bullet$).}
\label{fig:sketch_stability}
\end{figure}

To understand the formation and stability of the chiral structure, we test the three fundamental elastic deformations of the equilibrium nematic director field $\hat{d}$: splay $(\nabla\cdot\hat{d})^2$, bend $[\hat{d}\times(\nabla\times\hat{d})]^2$, and twist $[\hat{d}\cdot(\nabla\times\hat{d})]^2$ \cite{de_gennes_physics_1993}. 
For ideal splay, bend or twist deformations the total torque on a reference particle vanishes (Fig.~\ref{fig:sketch_stability}). 
However, the splay deformation is not stable in the SPP system because the system does not have sinks or sources. 
The bend deformation does not persist because there is no centrifugal force that would keep the particles on a curved path.
On the other hand, the SPP in an ideal twist deformation move within nematically ordered slices and therefore they persist.
Thus, due to symmetry, the twist pattern is the only one which is invariant under the motion of active particles.

\begin{figure*}
\centering
\includegraphics[width=\textwidth]{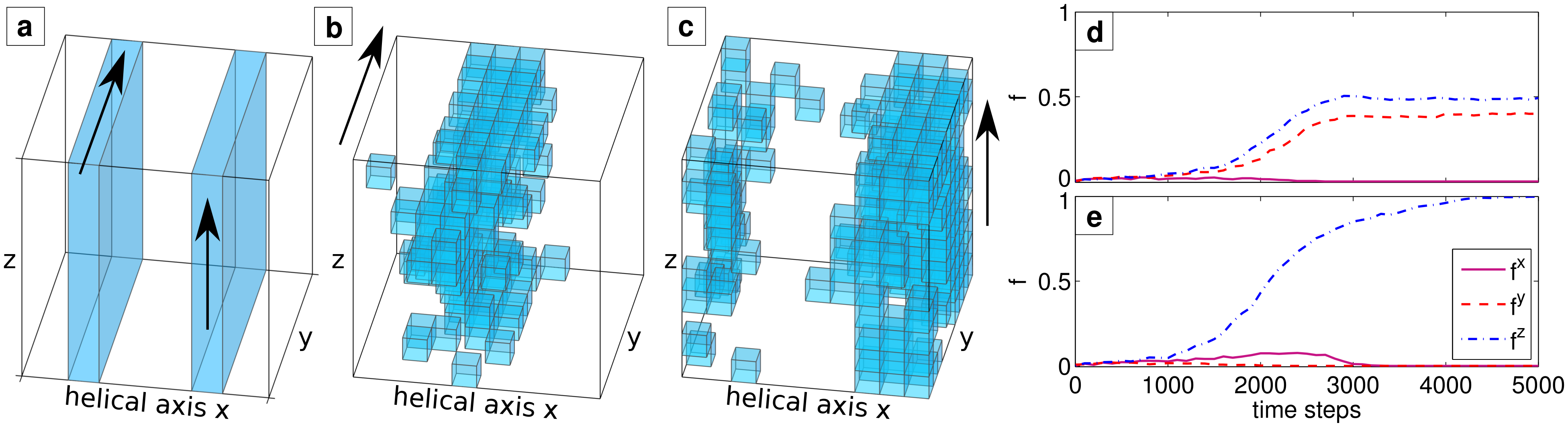}
\caption{(Color online) Formation of the chiral pattern. (a) Sketch of two competing layers (each nematically aligned). Here, the angle between the two $\hat{d}^\text{\,loc}$ is $\pi/2$. 
(b) and (c) Example from the same simulation as in Figs.~\ref{fig:snapshot} and \ref{fig:director_components} at time $t=2000$ time steps. The system is divided into $10^3$ boxes in each of which the fraction $f_{\text{box}}^{\{y,z\}}$ of particles with $\left|e_y\right|>0.9$ (b) and $\left|e_z\right|>0.9$ (c) is determined. Only boxes with $f_{\text{box}}^{\{y,z\}}>0.15$ are plotted. $f_{\text{box}}^{x}$ can be defined accordingly. (d) and (e) Evolution of the fraction $f^{\{x,y,z\}}$ of boxes with $f_{\text{box}}^{\{x,y,z\}}>0.15$ for the same chiral simulation (d) and a nematic example (e). The chiral structure shows two growing domains of similar size and different orientations, while only one orientation dominates in the nematic case.
}
\label{fig:boxes}
\end{figure*}

Next, we show how an achiral (nematic) interaction can lead to a chiral pattern by investigating the evolution of an isotropic to a chiral state in a simulation with large $\mathpzc{P}$. 
The interaction of the SPP leads to local alignment.
Firstly, areas of high local alignment grow with time. 
As these areas have grown to a certain size, they typically form a planar domain with $\hat{d}^\text{\,loc}$ perpendicular to the layer normal (Fig.~\ref{fig:boxes}(a)).
Secondly, in the evolution of the chiral state, we typically find two of these areas which fill almost the entire box and have a specific geometrical relation: the two planes have to be parallel and the two $\hat{d}^\text{\,loc}$ form an  angle close to $\pi/2$.
Domains with different $\hat{d}^\text{\,loc}$ start competing and that can result in a chiral structure.
Only long wavelength fluctuations can untwist a chiral configuration.
An example of the formation of these planes is shown in Fig.~\ref{fig:boxes}(b-c). 
It is quite simple to identify distinct and differently oriented domains that drive the formation of a chiral structure. 
We show the temporal evolution of domains with different orientations for a chiral structure (Fig.~\ref{fig:boxes}(d)) and also for a system that eventually evolves into a nematic state (Fig.~\ref{fig:boxes}(e)).

Seeding the system with two planar, ordered domains increases the probability of the chiral state (for a given $\mathpzc{P}$) from roughly few percent to about $50\%$.
This shows that the seeds are a good prerequisite for the chiral state but also that the fluctuations must play a key role since the probability does not reach $100\%$. 
Below, we elucidate the role of fluctuations in the driving mechanism.

The active motion of the particles stabilizes the chiral structure efficiently because orientational correlations propagate quickly. 
Notice, however, that  Eq.~\eqref{eq:differential_orientations} alone with vanishing noise can be seen as a Hamiltonian system that might show similar behavior, that is, one cannot exclude that in the appropriate circumstances orientational fluctuations can produce a chiral state also in an equilibrium system.

\begin{figure}
\includegraphics[width=\columnwidth]{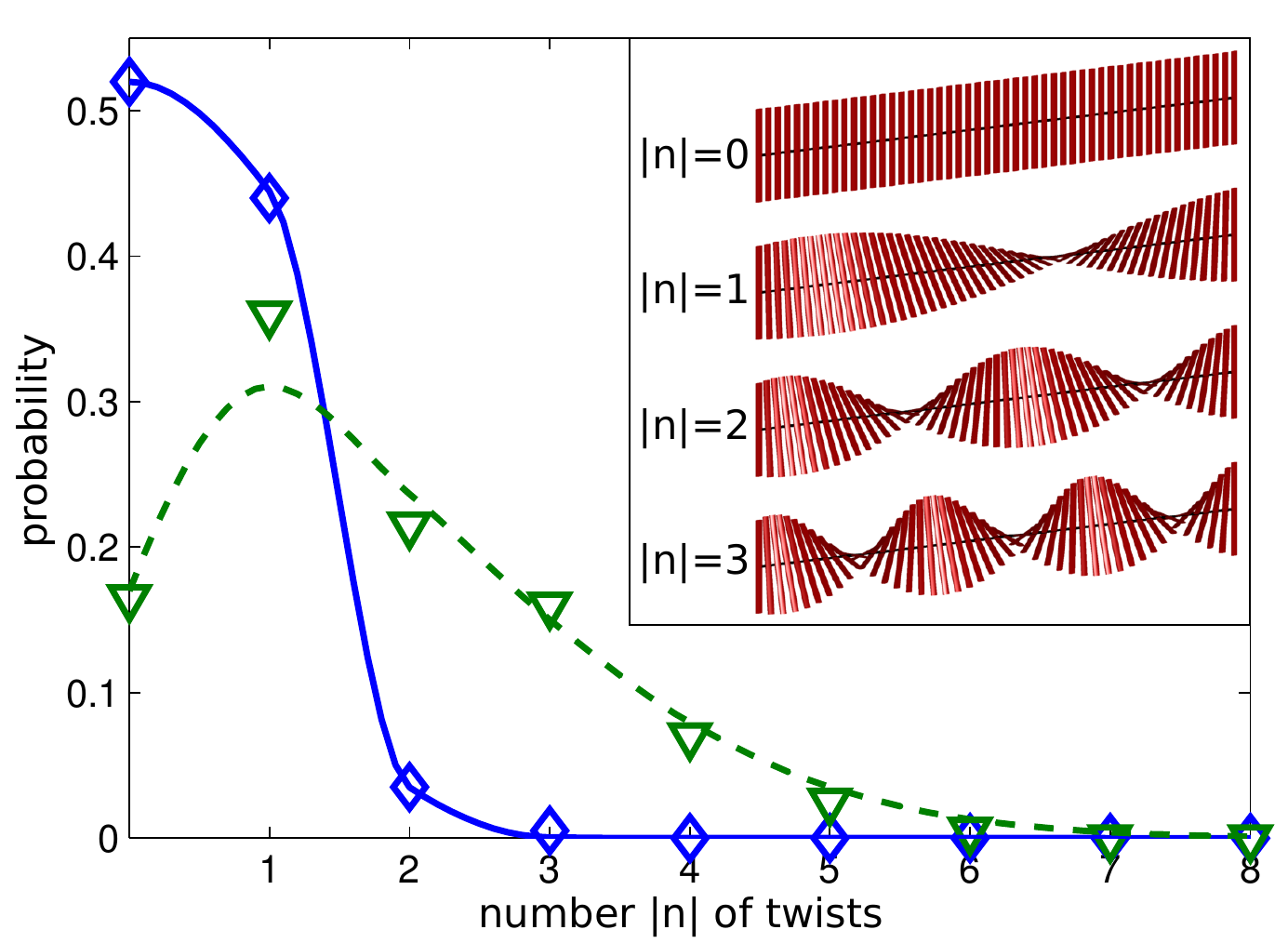}
\caption{(Color online) Inset: When annealed, the 1D nematic rotor model can evolve into the untwisted ground state or a chiral metastable state with $\pm n$ twists. Main panel: Frequency of final states has a maximum value at $\left|n\right|=1$ for a long chain with $N=800$ (green $\triangledown$, fit as green dashed line), and at $\left|n\right|=0$ for a short chain with $N=100$ (blue $\lozenge$, fit as blue solid line). 
The lines are guides for the eye.}
\label{fig:xymodel}
\end{figure}

Therefore, we consider a related, equilibrium statistical physics model equivalent to the one dimensional (1D) XY rotor model \cite{chaikin_principles_2000}: a 1D chain of $N$ nematic classical spins with Hamiltonian
\begin{equation}
H=\sum_{k=1}^{N}-J \cos\left[ 2\left(\theta_{k+1}-\theta_k\right)\right]+\frac{1}{2}I\left(\frac{\text{d} \theta_k}{\text{d}t}\right)^2
\label{eq:hamiltonian}
\end{equation}
where $\theta_k$  is the angle of spin $k$ with the $\hat{y}$-axis, rotating in the plane normal to the chain, and $I$ is the moment of inertia. 
The first term also represents the Lebwohl-Lasher interaction with prefactor $J\ge 0$ and the second term is the rotational kinetic energy. 
Periodic boundary conditions link the two ends of the chain. 
Our 1D system, having short range interactions, can have long range order only at $T=0$.
When a chain of finite length is annealed with $T\rightarrow0$, it either reaches the nematic ground state, or else falls into a chiral metastable state with a twist of $\pm n \pi$ (Fig.~\ref{fig:xymodel} inset), equivalently to a trapped spin wave in the XY model. 
To study the formation of such chiral states, we carried out dynamic simulations by integrating the equations of motion forward using the velocity Verlet method with a Langevin thermostat, annealing from  $T=10$ to $T=10^{-7}$ in $6\cdot 10^5$ time steps, with 200 annealing trials for each chain length.

Surprisingly, the most likely final state for chains with $N\geq 200$ is a metastable state with exactly one twist of either $\pm \pi$, while for shorter chains the most likely final state is untwisted, as shown in Fig.~\ref{fig:xymodel}.  
To explain this result, we consider the relative Boltzmann weight of each state, $\exp(-\Delta E_n/k_BT)$, where $\Delta E_n= 2J ({n \pi}/N)^2$ is the potential energy per spin of the system with $n$ twists \cite{Note_energy}.
As the exponent scales $\sim -n^2$, at any finite temperature for a chain of finite length, we expect a Gaussian distribution in $n$ of final states with a peak at $n=0$ and monotonically decreasing probability of finding a state with $n=\pm 1,\pm 2,\ldots$ twists. 
However, for long enough chains, the combined probability of the $n=\pm 1$ states exceeds that of the $n=0$ untwisted ground state and other values of $\left|n\right|$, explaining why the system must often land in a metastable state with exactly one twist in either direction. 

This result demonstrates that orientational fluctuations favor spontaneous formation of metastable chiral states, even in a system lacking intrinsic chirality at the microscopic level.  We also find that the mean square twist $\langle n^2\rangle$ increases linearly with chain length $N$.
 
A very similar behavior is found in the SPP model when a simulation box of aspect ratio $10:1:1$ is used ($N=10\cdot 23^3$, $\rho=1$, $\mathpzc{P}=0.08$). 
The probability of forming a chiral structure with twists of $\pm\pi$ is six times larger than for a cubic box.
Moreover, we also observe the spontaneous formation of structures with a twist of $\pm 2\pi$ in the elongated box.

In summary, we have shown that a model of SPP with achiral interactions can exhibit a spontaneous chiral symmetry breaking and that orientational fluctuations play a key role in the emergence of chiral structures. 
We have demonstrated that the twist deformation is the only stable configuration for a system of SPP. 
Moreover, the SPP model can be compared to an equilibrium statistical physics model of rotors finding a very similar chiral symmetry breaking.


\begin{thebibliography}{10}
\bibitem{Note_initial_conditions}
The formation of a chiral structure in the SPP model is not determined by the initial conditions: Starting from the same initial conditions (positions and directions of all particles), different realizations (different random seeds) can lead to a nematic, or a left- or right-handed chiral structure.
\bibitem{Note_normalization}
We note that the normalization depends on the radius (here: $L/4$) which defines the volume over which the local average (all $N_i$ particles) is taken.
The normalization is calculated analytically for an ideal chiral configuration with $\hat{e}=\pm(\cos(z\pi/L);\sin(z\pi/L);0)$.
\bibitem{Note_energy}
In $\Delta E_n$ the first term in Eq.~\eqref{eq:hamiltonian} is used: $\Delta E_n=\left<-J\cos(2\Delta\theta)\right>-E_0$ with $\Delta\theta=\frac{n\pi}{N}$ and $E_0=-J$. The small-angle approximation $\cos(2\Delta\theta)\approx 1-2\Delta\theta^2$ leads to $\Delta E_n=2J\left(\frac{n\pi}{N}\right)^2$.
\end{thebibliography}
\end{document}